# Graphene: A sub-nanometer trans-electrode membrane

S. Garaj[1], W. Hubbard[2], A. Reina[3], J. Kong[4], D. Branton[5] & J.A. Golovchenko[1,2*]

Submitted 12 April 2010 to *Nature*, where it is under review.

[1]*Department of Physics, Harvard University, Cambridge, MA 02138, USA.*

[2]*School of Engineering and Applied Sciences, Harvard University, Cambridge, MA 02138, USA.*

[3]*Department of Materials Science and Engineering, Massachusetts Institute of Technology, Cambridge, MA 02139, USA.*

[4]*Department of Electrical Engineering and Computer Science, Massachusetts Institute of Technology, Cambridge, MA 02139, USA.*

[5]*Department of Molecular and Cellular Biology, Harvard University, Cambridge, MA 02138, USA.*

*\*Corresponding Author*

**Isolated, atomically thin conducting membranes of graphite, called graphene, have recently been the subject of intense research with the hope that practical applications in fields ranging from electronics to energy science will emerge[1]. Here, we show that when immersed in ionic solution, a layer of graphene takes on new electrochemical properties that make it a trans-electrode. The trans-electrode's properties are the consequence of the atomic scale proximity of its two opposing liquid-solid interfaces together with graphene's well known in-plane conductivity. We show that several trans-electrode properties are revealed by ionic conductivity measurements on a CVD grown graphene membrane that separates two aqueous ionic solutions. Despite this membrane being only one to two atomic layers thick[2,3], we find it is a remarkable ionic insulator with a very small stable conductivity that depends on the ion species in solution. Electrical measurements on graphene membranes in which a single nanopore has been drilled show that the membrane's effective insulating thickness is less than one nanometer. This small effective thickness makes graphene an ideal substrate for very high-resolution, high throughput nanopore based single molecule detectors. Sensors based on modulation of graphene's in-plane electronic conductivity in response to trans-electrode environments and voltage biases will provide new insights into atomic processes at the electrode surfaces.**

We measured the trans-electrode ionic conductivity of a graphene membrane by mounting a 0.5 x 0.5 mm CVD grown sheet of graphene over a 200 x 200 nm aperture in 250 nm thick, free-standing, insulating $SiN_x$ layer on a silicon substrate frame (Fig. 1). Micro-Raman spectroscopy scans of the G, G' peaks from the membrane showed it to consist of 1 to 2 atomic layers of graphene[3,4]. The chip-mounted membrane was inserted in a fluidic cell such that it separated two compartments, each subsequently filled with ionic solutions in electrical contact with Ag/AgCl electrodes. Remarkably, the majority of membranes survived the pressure gradients and surface tension effects of this process.

With a 100 mV bias applied between the two Ag/AgCl electrodes, ionic current measurements show that the graphene membrane's trans-ionic conductance is far below the nS level (Table I). The highest conductivities are observed for solutions with the largest atomic size cations, Cs and Rb, probably a reflection of their minimal hydration shell that mediates their interaction with the graphene surface[5,6]. This conductivity may be ion transport through the graphene or faradaic. Small asymmetries and nonlinearities in the I-V curves were observed in the measurements from which the data in Table 1 and elsewhere (e.g., Fig. 2) were obtained, likely due to double layer phenomena near biased membranes[7], or to asymmetrical absorption of impurities on the graphene surfaces during CVD growth[3].

Drilling a single nanometer scale pore[8] in the graphene trans-electrode membrane increases its conductivity by orders of magnitude (Fig. 2) and enables experiments to evaluate the graphene electrode's effective insulating thickness. Because the nanopore diameter, the solution conductivity, and the membrane's insulating thickness control trans-electrode conductivity, experiments with known nanopore diameters and solution conductivities allow one to deduce graphene's effective insulating thickness.

The ionic conductivity $G$ from a pore of diameter $d$ in an infinitely thin insulating membrane is given by[9]

$$G_{thin} = \sigma \cdot d \tag{1}$$

where $\sigma = F(\mu_{K^+} + \mu_{Cl})c$ is the conductivity of the ionic solution, F is the Faraday constant, c is ionic concentration, and $\mu_i(c)$ is the mobility of potassium ($i = K$) and chloride ($i = Cl$) ions used in our measurements. The current density in the case of a nanopore in a very thin membrane is sharply peaked at the pore's perimeter. Thus, when the membrane is much thinner than the pore diameter the conductivity becomes proportional to the pore diameter, rather than its area[10]. For finite but small thicknesses we rely on computer simulations to predict the conductivities.

We prepared pores with diameters ranging from 5 to 23 nanometers (as determined by TEM) and measured their trans-electrode conductivities (Fig. 3). In agreement with equation (1), a near linear dependence on pore diameter is clearly observed. Fig. 3 also shows the results of computer calculations of nanopore conductivity in an idealized uncharged, insulating membrane, as a function of pore diameter and membrane thickness. These calculated results are obtained by numerically solving the relevant Laplace equation for the ionic current density, with appropriate solution conductivity and boundary conditions, and then integrating this over the pore area to get the conductivity[11]. We refer to this thickness $L$ used in this idealized model as the Graphene Insulating Thickness, or $L_{GIT}$, to distinguish this important phenomenological electrical property of the trans-electrode membrane system. The best fit to the measured pore conductance data in Fig. 3 yields $L_{GIT} = 0.6$ nm, with an uncertainty of -0.6 and + 0.9 nm. Fig. 3 also shows the calculated results for $L_{GIT} = 2.0$ nm and $L_{GIT} = 10.0$ nm.

Measurements of nanopore conductivity while it is being traversed by an insulating long chain polymer of DNA provide an alternative method of evaluating $L_{GIT}$. In such experiments, the negatively charged DNA molecules are electrophoretically drawn to and driven through a nanopore. As each insulating molecule passes through the pore, it generates an "event" that transiently reduces, or blocks, the ionic conductivity in a manner that reflects both polymer size and conformation[12]. As we will show, DNA experiments can also reveal the membrane thickness and the nanopore diameter. The results using a 5 nm pore in graphene and double stranded DNA molecules are shown in Fig. 4. The insets show two single molecule translocation events. In the right hand event the molecule passes through the pore in an orderly unfolded linear fashion, and in the left hand event the molecule is folded over on itself when it enters the pore, increasing the current blockade for a short time[12]. Each single molecule translocation event can be characterized by two parameters: the average current drop, or blockade, and the duration of the blockade, which is the time it takes for the molecule to completely translocate through the pore. The scatter plot in Fig. 3 shows the value of these parameters for each of 400 DNA single molecule events. The characteristic shape of this data is similar to that obtained in silicon nitride nanopore experiments[12]. Almost all the events, folded and unfolded, fall near a line of constant electronic charge deficit (*ecd*)[12]. This indicates that these molecules' passage through the pore is not inhibited by sticking to the graphene surface. Those few events that are encircled in the plot do not satisfy this condition and their long translocation times are likely due to graphene-DNA interactions, which slow their translocation through the nanopore.

We compare the experimentally determined open pore and DNA blocked pore conductivities with numerical calculations where the membrane thickness and the nanopore diameter are the fitting parameters. Using the observed mean current blockade $\Delta I = 1.24 \pm 0.08$ nA during translocation of single file double stranded DNA, $d = 2.0$ nm[13], and the observed conductance of the pore $G = 105 \pm 1$ nS absent DNA (data in Fig. 3), we calculate that $L_{GIT} = 0.6 \pm 0.5$ nm, in excellent agreement with the value deduced above from open pore measurements. The pore diameter $d_{GIT} = 4.6 \pm 0.4$ nm deduced from these calculations also agrees with the geometric diameter of $5 \pm 0.5$ nm obtained from a TEM of this pore.

The best fit value $L_{GIT} = 0.6$ nm from both experiments agrees with molecular dynamics simulations showing the graphene-water distance to be 0.31-0.34 nm on each side of the membrane[14,15]. $L_{GIT}$ might also be influenced by the presence of immobilized water molecules and adsorbed ions in the Stern layer[7]. On the other hand, theoretical and experimental studies suggest an anomalously strong slip between water and internal carbon nanotube surfaces[14,16], arguing against the presence of any immobilized water layer. Although very little is actually known about the surface chemistry of specifically adsorbed ions on single graphene layers[1], measurements of the ionic current through the inner volume of carbon nanotubes with diameters less than 1nm[17] indicate that ions may not be immobilized at all on the graphitic surfaces. Our sub-nanometer values for $L_{GIT}$ support this view.

The extremely small $L_{GIT}$ value we obtain suggests that nanopores in graphene membranes are uniquely optimal for discerning spatial or chemical molecular structure along the length of molecule as it passes through the pore. Although polymer translocation speeds and electronics bandwidth currently preclude a direct measurement of a nanopore's spatial or geometric resolution limit[18], we can gain insight into the system's limit by numerically modeling the resolution obtainable as a function of $L_{GIT}$.

The model assumes a long insulating 2.2 nm diameter cylinder symmetrically translocating through the center of a 2.4 nm diameter nanopore. At one position along its length, the cylinder diameter changes discontinuously from 2.2 nm to 2.0 nm. Solving the conductivity for this geometry as the discontinuity passes through the pore, we obtain the predictions shown in Fig. 5. The decreasing blockade (increasing conductivity) of a pore is clearly seen as the large diameter portion of the cylinder exits the pore. The results of calculations for two $L_{GIT}$ values are shown. For the conservative $L_{GIT}$ = 1.5 nm, the spatial resolution (defined as the distance over which the conductivity changes from 75% of its greatest value to 25% of that value) is given by $\delta z_{GIT} = 7.5$ Å, whereas the best-fit value $L_{GIT} = 0.6$ nm leads to $\delta z_{GIT} = 3.5$ Å. We conclude from our experiments and modeling that a pore in graphene is inherently capable of probing molecules with subnanometer resolution. Functionalizing the graphene nanopore boundary[5] or observing its local in-plane electrical conductivity during translocations may provide additional or alternative means of further increasing the resolution of this system.

Finally, we note that when immersed in an ionic solution, a single atomic layer of graphene takes on new surface electrochemical capabilities that make it a trans-electrode: an electrode whose two opposing surface-liquid interfaces are in atomic scale proximity. As such, the structure is a single entity with unique and useful properties that make it a particularly interesting device for sensors and surface electrochemistry studies. The presence of external fields due to applied solution bias and the proximity and dynamics of solution ions near both graphene surfaces will affect both the mobility and concentration of graphene's in-plane charge carriers, providing a microscopic electronic view of surface properties analogous to those that have recently been revealed in electronic nanopore surface studies[19]. Many opportunities exist for modifying a graphene trans-electrode by chemical doping or by introducing vacancy type defects. The interactions at, and between, the two liquid-solid interfaces in graphene may well hold many surprises.

**Acknowledgements**  This work was funded by a grant to J.A.G. and D.B. from the National Human Genome Research Institute, National Institutes of Health.

**Author Contributions**  S.G. conceived and performed most of the experiments.  W.H., A.R. and J.K. prepared and characterized the graphene.  S.G., D.B. and J.A.G. drafted the manuscript.  All authors discussed the results and commented on the manuscript.

**Author Information**  Correspondence and requests for materials should be addressed to S.G. (sgaraj@fas.harvard.edu).

**Table I**

| Solution | Graphene Conductance (pS) | Sol. Conductivity ($10^{-3} Sm^{-1}$) | Hydration energy[20] (eV) |
|---|---|---|---|
| CsCl | 67±2 | 1.42 | 3.1 |
| RbCl | 70±3 | 1.42 | 3.4 |
| KCl | 64±2 | 1.36 | 3.7 |
| NaCl | 42± 2 | 1.19 | 4.6 |
| LiCl | 27±3 | 0.95 | 5.7 |

**Table Legend.** Trans-membrane conductivity of an as-grown graphene membrane separating two compartments each containing the ionic solutions indicated in column 1. Conductivities were determined from voltage bias scans between +100 *mV* and -100 *mV.* All data shown here are from the same membrane. The absolute magnitudes of the conductivities varied by a factor of two from membrane to membrane, but the systematic variation with ionic solutions was invariant for all membranes.

**Figure Legends**

**Figure 1. Schematic of our experiments.** A graphene membrane was mounted over a 200 x 200 nm aperture in $SiN_x$ suspended across a Si frame. The membrane separates two ionic solutions in contact with Ag/AgCl electrodes. *Inset:* A graphene membrane into which a nanopore has been drilled.

**Figure 2. Trans-electrode I–V curves.** Results for an as-grown graphene membrane (dashed line) and a membrane with a 5 nm pore (solid line) are shown. The ionic solutions were unbuffered 1M KCl solution, pH ≈ 5.2. The ionic conductivity of the pore is quantitatively in agreement with the modeling presented in the text. Applying bias voltages in excess of ~250 mV gradually degraded the insulating properties of the membranes. Insets, *top*: transmisson electron micrograph (TEM) of a mounted graphene membrane; *bottom*: a TEM images of the 5 nm pore.

**Figure 3. Graphene nanopore conductivity.** Closed circles are experimental results a 1 M KCl solution of conductivity $\sigma = 11$ S/m$^{-1}$. The solid curve shows the modeled conductivity of a 0.6 nm insulating membrane and is the best fit to the measured conductivities. Modeled conductivities for a 2 nm thick membrane (dotted line) and a 10 nm thick membrane (dash-dot line) are presented for comparison.

**Figure 4. Average nanopore current blockades vs. blockade duration.** A 10 kbp fragment of λ DNA (16 µg/ml) was electrophoretically driven through a 5 nm diameter graphene pore by an applied voltage bias of 160 mV. The graphene membrane separated two fluid cells containing unbuffered 3M KCl solutions, pH 10.4. Insets show typical current-time traces for two translocation events sampled from among those pointed to by the arrows. The hyperbolic curve corresponds to freely translocating events at a fixed *ecd*[12]. Encircled events are delayed by graphene DNA interactions.

**Figure 5. Geometric Resolution.** Modeled nanopore conductivity as the abrupt diameter decrease of a model molecule (inset) translocates through a 2.4 nm pore. The attainable resolution for two membranes of different insulating thicknesses is assumed to be achieved when the measured current through the nanopore changes from 75% to 25% of the maximum blockade change that would occur as the model molecule translocates through the nanopore.

**FIGURES**

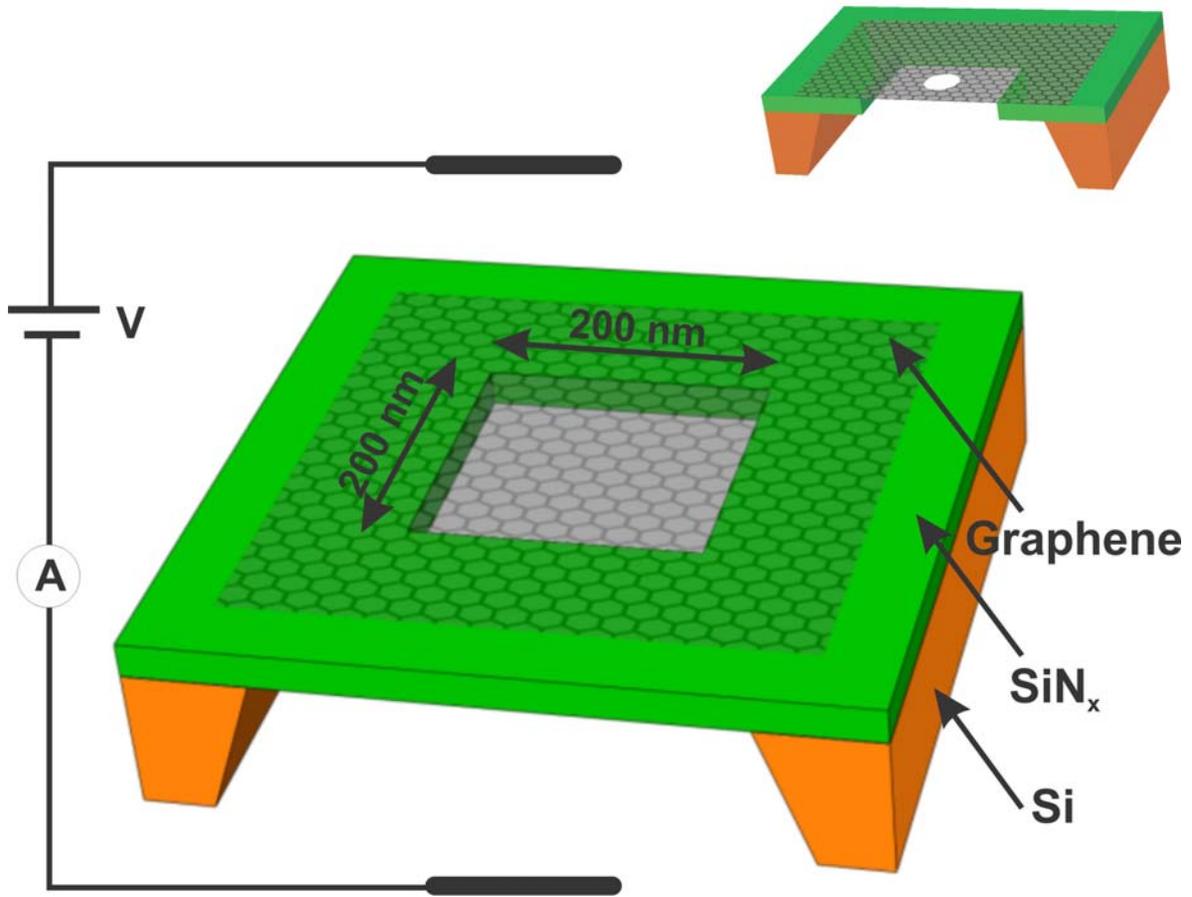

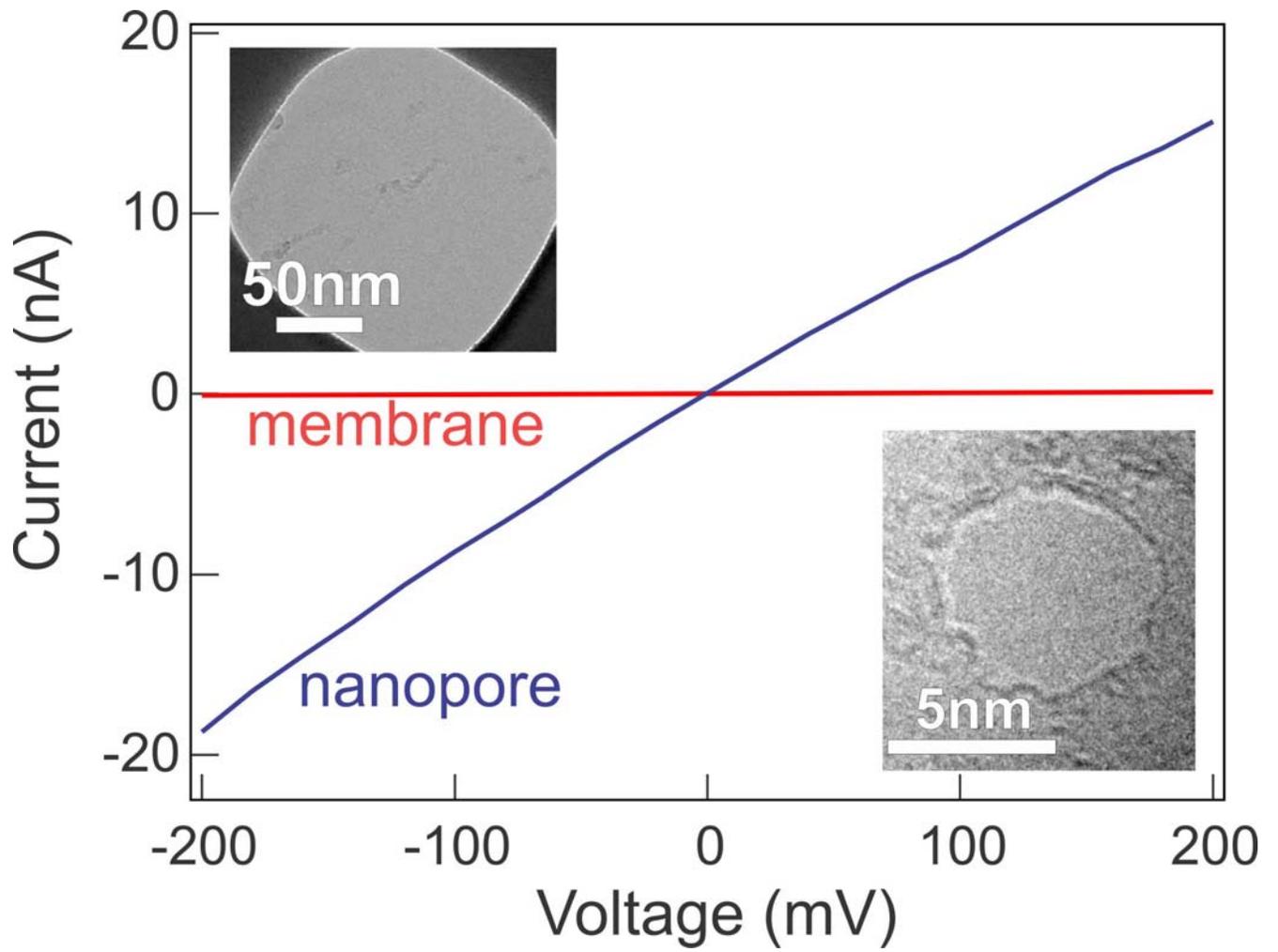

**Figure 2**

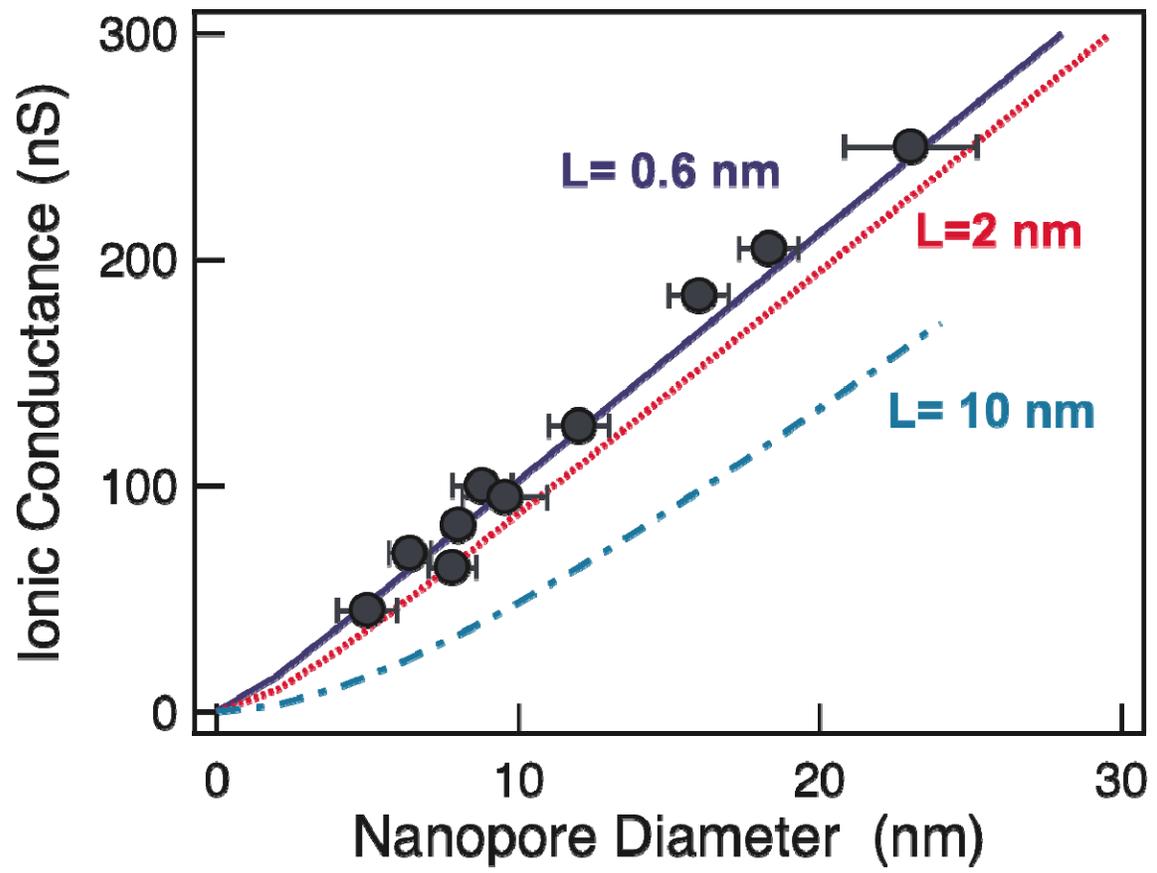

**Figure 3**

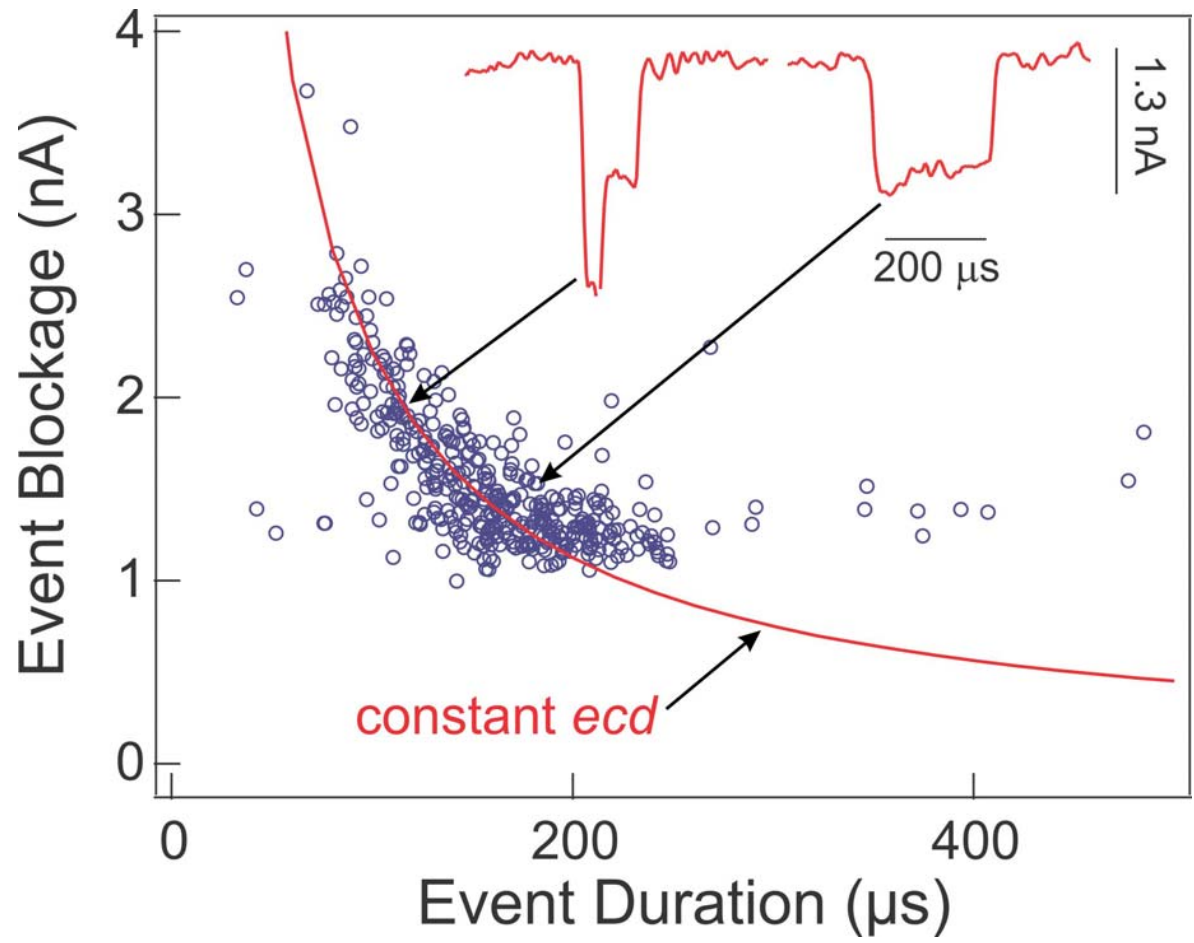

**Figure 4**

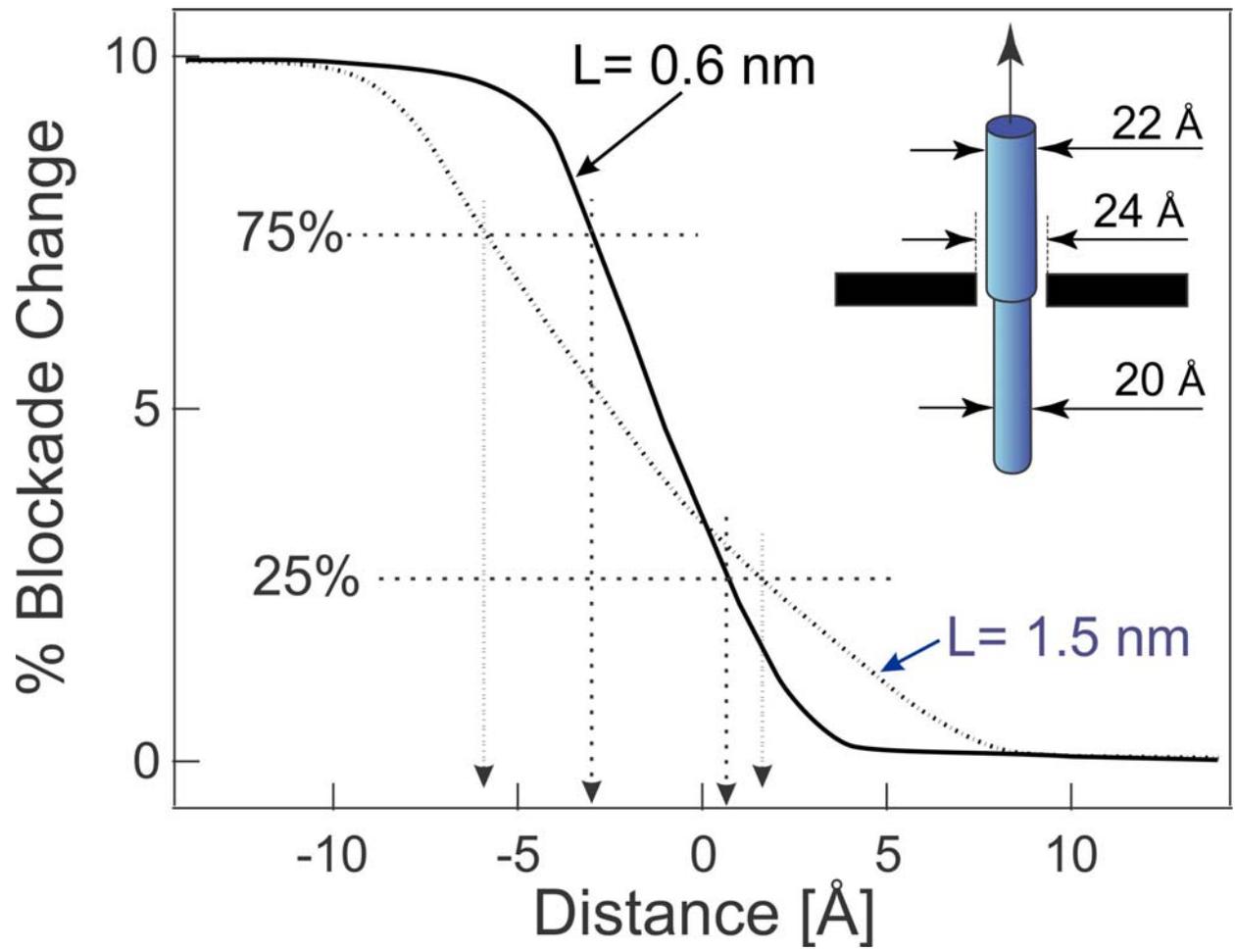

**Figure 5**